\documentclass[prd,aps,showpacs,tightenlines,nofootinbib,twocolumn]{revtex4}
\usepackage{graphicx}

\newcommand{\CL}{{\cal L}}

\newcommand{\CO}{{\cal O}}

\newcommand{\bear}{\begin{array}}  \newcommand{\eear}{\end{array}}
\newcommand{\bea}{\begin{eqnarray}}  \newcommand{\eea}{\end{eqnarray}}
\newcommand{\beq}{\begin{equation}}  \newcommand{\eeq}{\end{equation}}
\newcommand{\bef}{\begin{figure}}  \newcommand{\eef}{\end{figure}}
\newcommand{\bec}{\begin{center}}  \newcommand{\eec}{\end{center}}
\newcommand{\non}{\nonumber}  
\newcommand{\lmk}{\left(}  \newcommand{\rmk}{\right)}
\newcommand{\lkk}{\left[}  \newcommand{\rkk}{\right]}
\newcommand{\lhk}{\left \{ }  \newcommand{\rhk}{\right \} }
\newcommand{\del}{\partial}  

\newcommand{\bib}{\bibitem}

\def\IB#1#2#3{{\bf #1}, #2 (19#3)}
\def\IBB#1#2#3{{\bf #1}, #2 (20#3)}
\def\IBID#1#2#3{{\it ibid}. {\bf #1}, #2 (19#3)}
\def\IBIDD#1#2#3{{\it ibid}. {\bf #1}, #2 (20#3)}

\def\AAA#1#2#3{Astron. Astrophys. {\bf #1}, #2 (20#3)}

\def\APJJ#1#2#3{Astrophys. J. {\bf #1}, #2 (20#3)}
\def\APJL#1#2#3{Astrophys. J. Lett. {\bf #1}, L#2 (19#3)}
\def\APJLL#1#2#3{Astrophys. J. Lett. {\bf #1}, L#2 (20#3)}

\def\GRG#1#2#3{Gen. Relativ. Gravit. {\bf #1}, #2 (19#3)}

\def\JHEP#1#2#3{J. High Energy Phys. {\bf #1}, #2 (19#3)}

\def\MNRAS#1#2#3{Mon. Not. R. Astron. Soc. {\bf #1}, #2 (19#3)}

\def\NATT#1#2#3{Nature (London) {\bf #1}, #2 (20#3)}
\def\NPB#1#2#3{Nucl. Phys. {\bf B#1}, #2 (19#3)}
\def\NPBB#1#2#3{Nucl. Phys. {\bf B#1}, #2 (20#3)}

\def\PLB#1#2#3{Phys. Lett. B {\bf #1}, #2 (19#3)}
\def\PLBB#1#2#3{Phys. Lett. B {\bf #1}, #2 (20#3)}
\def\PLBold#1#2#3{Phys. Lett. {\bf#1B}, #2 (19#3)}

\def\PRD#1#2#3{Phys. Rev. D {\bf #1}, #2 (19#3)}
\def\PRDD#1#2#3{Phys. Rev. D {\bf #1}, #2 (20#3)}

\def\PRL#1#2#3{Phys. Rev. Lett. {\bf#1}, #2 (19#3)}
\def\PRLL#1#2#3{Phys. Rev. Lett. {\bf#1}, #2 (20#3)}

\def\PRTT#1#2#3{Phys. Rep. {\bf#1}, #2 (20#3)}
\def\PTP#1#2#3{Prog. Theor. Phys. {\bf #1}, #2 (19#3)}

\def\RPP#1#2#3{Rep. Prog. Phys. {\bf #1}, #2 (19#3)}

\def\RMP#1#2#3{Rev. Mod. Phys. {\bf #1}, #2 (19#3)}

\newcommand{\gtrsim}{ \mathop{}_{\textstyle \sim}^{\textstyle >} }
\newcommand{\lesssim}{ \mathop{}_{\textstyle \sim}^{\textstyle <} }

\begin{document}


\title{Generation of cosmological large lepton asymmetry from a
  rolling scalar field}
\author{Masahide Yamaguchi}
\affiliation{Research Center for the Early Universe, University of Tokyo,
Tokyo 113-0033, Japan \\ and \\
Physics Department, Brown University, Providence, Rhode Island 02912, USA}
\date{\today}
\begin{abstract}
  We propose a new scenario to simultaneously explain a large lepton
  asymmetry and a small baryon asymmetry. We consider a rolling scalar
  field and its derivative coupling to the lepton number current. The
  presence of an effective nonzero time derivative of the scalar
  field leads to $CPT$ violation so that the lepton asymmetry can be
  generated even in thermal equilibrium as pointed out by Cohen and
  Kaplan. In this model, the lepton asymmetry varies with time.  In
  particular, we consider the case where it grows with time. The final
  lepton asymmetry is determined by the decoupling of the lepton
  number violating interaction and can be as large as order unity. On
  the other hand, if the decoupling takes place after the electroweak
  phase transition, a small baryon asymmetry is obtained from the
  small lepton asymmetry at that time through sphaleron effects.  We
  construct a model in which a rolling scalar field is identified
  with a quintessence field.
\end{abstract}

\pacs{98.80.Cq \hspace{8cm} BROWN-HET-1333}
\maketitle


\section{Introduction}

\label{sec:introduction}

The baryon number density dominates over the antibaryon number density
in our Universe. The magnitude of the baryon asymmetry is mainly
estimated by the two different methods. One is big bang
nucleosynthesis (BBN). By comparing predicted primordial abundances of
light elements (D, $^3$He, $^4$He and $^7$Li) with those inferred from
observations, the baryon-to-entropy ratio $n_B/s$ is estimated as
$n_B/s \sim (8-10) \times 10^{-11}$ \cite{BBN,BBN2}. The other is
observations of small scale anisotropies of the cosmic microwave
background (CMB) radiation
\cite{BOOMERANG,MAXIMA,DASI,CBI,Archeops,BM,WMAP}.  The
baryon-to-entropy ratio inferred from recent results of the Wilkinson
Microwave Anisotropy Probe (WMAP) roughly coincides with that inferred
from BBN. However, the detailed analysis shows that the best fit value
of the effective number of neutrino species $N_{\nu}$ is significantly
smaller than $3.0$ \cite{BBN3}. Of course, $N_{\nu} = 3.0$ is
consistent at $\sim 2 \sigma$ and such discrepancies may be completely
removed after observations are improved and their errors are reduced.
However it is probable that such small discrepancies are genuine and
suggest additional physics in BBN and the CMB.

An interesting possibility to eliminate such discrepancies is the
presence of a large and positive lepton asymmetry of electron type
\cite{KS}. Roughly speaking, such an asymmetry causes two effects on
the predicted primordial abundance of $^{4}$He. The excess of electron
neutrinos shifts the chemical equilibrium between protons and neutrons
toward protons, which reduces the predicted primordial abundance of
$^{4}$He. On the other hand, the excess of electron neutrinos also
causes increase of the Hubble expansion, which makes the predicted
primordial abundance of $^{4}$He increase. In practice, the former
effect overwhelms the latter so that the presence of a large and
positive electron type asymmetry decreases the predicted primordial
abundance of $^{4}$He, which often solves the discrepancies as
mentioned above. Furthermore, large lepton asymmetries are useful to
realize the cool dark matter model of the large scale structure
formation \cite{cool} and the relic neutrino scenario of the extremely
high energy cosmic rays \cite{GK}. However, it is very difficult for
such a large lepton asymmetry to be compatible with a small baryon
asymmetry. This is mainly because, if we take the sphaleron effects
into account, lepton asymmetry is converted into baryon asymmetry of
the same order with the opposite sign \cite{sphaleron}.

There are several ways to overcome this difficulty. One possibility is
to generate the large lepton asymmetry after electroweak phase
transition but before BBN, which may be realized through oscillations
between active neutrinos and sterile neutrinos \cite{oscillation}.
Another is to disable the sphaleron effects. It was pointed out that the
presence of a large chemical potential prevents restoration of the
electroweak symmetry \cite{nonrestoration}. Based on this nonrestoration
mechanism, the generation of a large lepton asymmetry compatible with
the small baryon asymmetry was discussed \cite{second}.  March-Russell
{\it et al.} discussed another possibility \cite{MRM}.  In their model,
a positive electron type asymmetry but no total lepton asymmetry, that
is, $L_{e} = - L_{\mu} > 0$ and $L_{\tau} = 0$, is generated by the
Affleck-Dine mechanism for some flat direction \cite{AD}. Then, the
small positive baryon asymmetry is produced through thermal mass effects
of sphaleron processes.  Recently, Kawasaki, Takahashi, and the present
author discussed another possibility \cite{KTY}, in which a positive
electron type asymmetry but negative total lepton asymmetry is produced
by the Affleck-Dine mechanism and almost all the produced lepton numbers
are absorbed into L-balls. A small amount of negative lepton charges is
evaporated from the L-balls due to thermal effects. These are converted
into the observed small baryon asymmetry by virtue of sphaleron
effects. However, the remaining lepton charges are protected from
sphaleron effects and released into thermal plasmas by the decay of
L-balls before BBN.

In this paper, we discuss another possibility, in which a spontaneous
leptogenesis mechanism proposed by Cohen and Kaplan is used \cite{CK}.
We consider a real scalar field, which slow rolls like a quintessence
field and couples derivatively to the lepton number current. The
presence of an effective nonzero time derivative of the scalar field
leads to $CPT$ violation so that the lepton asymmetry can be generated
even in thermal equilibrium \cite{DZ}. The produced lepton asymmetry
varies with time according to the dynamics of the rolling scalar
field. Then, the final lepton asymmetry is determined by the decoupling
of the lepton number violating interaction and it can be as large as
order unity. On the other hand, in the case that the lepton asymmetry
grows with time, it can be small at the electroweak phase transition,
which is converted into the observed small baryon asymmetry. In a
similar way to \cite{KTY}, we can generate a positive electron type
asymmetry but a negative total lepton asymmetry according to the
coupling constants. Thus, a large positive electron type asymmetry and
the small positive baryon asymmetry are realized simultaneously.

In the next section, after reviewing the spontaneous baryo/leptogenesis
mechanism proposed by Cohen and Kaplan, we explain our scenario to
produce the desired asymmetries. In Sec. III, we give a model as an
example, in which a real rolling scalar field is identified with a
quintessence field. Several authors have already discussed similar
scenarios, in which only the present small baryon asymmetry is explained
by a Nambu-Goldstone boson \cite{DF}, a quintessence field
\cite{LFZ,FNT}, and an inflaton in warm inflation \cite{RY}, with their
derivative couplings to a baryon current or a lepton current.

\section{Spontaneous leptogenesis and large lepton asymmetry}

\label{sec:spontaneous}

First of all, we introduce the derivative coupling of a real rolling
scalar field $\phi$ with the lepton number current $J_{i}^{\mu}$,
\beq
  \CL_{\rm eff} = \sum_{i} \frac{c_{i}}{M}\,f(\phi)\, 
                    \del_{\mu}\phi\,J_{i}^{\mu},
  \label{eq:coupling}
\eeq
where $i$ denotes the generation, $c_{i}$ are coupling constants,
$f(\phi)$ is a function of $\phi$, and $M$ is the cutoff scale, which
is set to be the reduced Plank mass $M_{G} \sim 10^{18}$ GeV in this
paper. Assuming that $\phi$ is homogeneous,
\beq
  \CL_{\rm eff} = \sum_{i} \frac{c_{i}}{M}\,f(\phi)\, 
                    \dot\phi\,n_{L}^{i}
                = \sum_{i} \mu_{i}(t)\,n_{L}^{i},
\eeq
where the effective time-dependent chemical potential $\mu_{i}(t)$ is
given by
\beq
  \mu_{i}(t) \equiv \frac{c_{i}}{M} f(\phi)\, \dot\phi.
\eeq
As pointed out by Cohen and Kaplan, this interaction induces $CPT$
violation if the time derivative $\dot\phi$ is effectively nonzero,
which generates the lepton asymmetry even in a state of thermal
equilibrium. Then, in thermal equilibrium, the lepton asymmetry
$n_{L}^{i}$, for $\mu_{i} < T$, is given by
\beq
  n_{L}^{i} = \frac{g}{6} T^3 \lhk 
                  \frac{\mu_{i}}{T} 
                     + \CO\lkk\lmk\frac{\mu_{i}}{T}\rmk^3\rkk
                                \rhk,
\eeq
where $g$ represents the number of degrees of freedom of the fields
corresponding to $n_{L}$. Since the entropy density $s$ is given by $s
= \frac{2\pi}{45}g_{\ast}T^3$ with $g_{\ast}$ the total number of
degrees of freedom for the relativistic particles, the ratio between
the lepton number density and the entropy density is given by
\beq
  \frac{n_{L}^{i}}{s} \simeq \frac{15}{4\pi^2}\frac{g}{g_{\ast}}
                                \frac{\mu_{i}}{T}
                        = \frac{15}{4\pi^2}\frac{g}{g_{\ast}}
                           \frac{c_{i}}{MT} f(\phi)\, \dot\phi.
  \label{eq:lepton}
\eeq

The above generation mechanism is effective as long as the lepton number
violating interaction is in thermal equilibrium. Thus, the final lepton
asymmetry is determined by the decoupling temperature $T_{D}$ of the
lepton number violating interaction. Such a lepton number violating
interaction with the low decoupling temperature can be obtained in the
context of the Zee model \cite{Zee} or the triplet Higgs boson with its
lepton number not equal to two. However, we do not specify the lepton
number violating interaction in this paper. Instead, for our purpose, we
have only to demand that the decoupling of the lepton number violating
interaction takes place after the electroweak phase transition.

If $c_1$ is positive but the total sum of $c_{i}$ is negative,
positive electron type asymmetry but negative total lepton asymmetry
is realized. Depending on the values of the scalar field $\phi$ and
its time derivative $\dot\phi$, the absolute magnitude of the lepton
asymmetry can be as large as order unity. On the other hand, the
lepton asymmetry at the electroweak phase transition can be as small
as the order of $10^{-10}$ according to the dynamics of the scalar
field. Then, a part of the lepton asymmetry at that time is converted
into the baryon asymmetry through the sphaleron processes, which can
be estimated as \cite{sphaleron}
\bea
  \eta = \frac{n_B}{s} &\simeq& \left. 
                  - \frac{8}{23} \sum_{i} \frac{n_{L}^{i}}{s}\,
                  \right|_{T_{\rm EW}} \non \\    
       &=& - c \left.  
                    \frac{30}{23\pi^2}\frac{g}{g_{\ast}}
                           \frac{1}{MT} f(\phi)\, \dot\phi\,
                  \right|_{T_{\rm EW}},      
\eea
where $c = \sum_{i} c_{i}$, $T_{\rm EW}$ represents the temperature at
the electroweak phase transition, and we have assumed the standard
model with two Higgs doublets and three generations. Thus, the large
positive electron type asymmetry and the small positive baryon
asymmetry are obtained.

\section{Quintessential baryo/leptogenesis model}

\label{sec:quintessential}

In this section, as an example, we consider the case in which the real
scalar field is identified with a quintessence field. Though a lot
of quintessence models are proposed
\cite{RP,FHW,CSN,TW,CDS,BS,AS,COY,AMS}, we take a k-essence model
\cite{COY,AMS}, in which the quintessence field can naturally have
the derivative coupling to the lepton current.

One of the candidates to realize the k-essence model is the gauge-neutral
massless scalar fields present in string theory. The low-energy
effective action $S_{\rm eff}$ has the following form \cite{DP}:
\bea
  S_{\rm eff} &=& \frac{1}{6\kappa^{2}}
                   \int d^{4}x \sqrt{\hat{g}}
                    \lhk 
         - B_{g}(\phi)\hat{R} - B_{\phi}^{(0)}(\phi)(\hat{\nabla}\phi)^{2}
                    \right.
         \non \\   
         &+&         \left. 
           \alpha'
                    \lkk 
         c_{1}B_{\phi}^{(1)}(\phi)(\hat{\nabla}\phi)^{4}
         + \sum_{i} c_{2}^{(i)} \overline{\psi_{i}}\hat{D}\psi_{i}
                    \rkk
         + \cdots
                    \rhk, 
\eea
where $\kappa = 8\pi G/3$, $\phi$ is the dilaton or the moduli, and
$\psi_{i}$ are the leptons, respectively. $B_{j}(\phi)$ are the
coupling functions and could take the complicated forms in the strong
coupling regime. In the Einstein frame where $g_{\mu\nu} =
B_{g}(\phi)\hat{g}_{\mu\nu}$, the effective action becomes
\beq
  S_{\rm eff} = \int d^{4}x \sqrt{g}
                    \lkk 
                -\frac{R}{6\kappa^{2}} + p(\phi,\nabla\phi)
                    \rkk
\eeq
with
\beq
  p(\phi,\nabla\phi) = p_{\phi}(\phi,\nabla\phi) + p_{d}(\phi,\nabla\phi).
\eeq
The function $p_{\phi}(\phi,\nabla\phi)$ depends on only $\phi$ and $X
\equiv (\nabla\phi)^2 /2$, and is given by
\beq
  p_{\phi}(\phi,X) = K(\phi) X + L(\phi) X^2
\eeq
with 
\bea
  K(\phi) &=& \frac{1}{6\kappa^{2}}    
                \lkk
                3 \frac{B_{g}'^{2}(\phi)}{B_{g}^{2}(\phi)}
                - 2 \frac{B_{\phi}^{(0)}(\phi)}{B_{g}(\phi)} 
                \rkk ,\non \\
  L(\phi) &=& \frac{2\alpha'}{3\kappa^2} c_{1} B_{\phi}^{(1)}(\phi).
\eea
On the other hand, the function $p_{d}(\phi,\nabla\phi)$ is given by
\beq
  p_{d}(\phi,\nabla\phi) = \sum_{i} M_{i}(\phi) \del_{\mu}\phi J_{i}^{\mu}
\eeq
with
\beq
  M_{i}(\phi) = \frac{\alpha'}{4\kappa^2} c_{2}^{(i)} 
                 \frac{B_{g}'(\phi)}{B_{g}^{\frac52}(\phi)}.
\eeq
The above derivative coupling to the lepton current is obtained after
integration by parts and the prime denotes $d/d\phi$. 

More phenomenologically, the above Lagrangian can be obtained by
imposing a real scalar field $\phi$ on the shift symmetry as proposed
in \cite{KYY,YY},
\beq
  \phi \rightarrow \phi + C M_{G},
\eeq
with $C$ is a dimensionless parameter. Then, the Lagrangian depends on
only $\del_{\mu} \phi$ so that $\phi$ can naturally have the
derivative coupling to the lepton current. The functions $K(\phi)$,
$L(\phi)$, and $M_{i}(\phi)$ may be associated with the breaking of
such a symmetry.

After redefining the scalar field $\phi$ such that
\beq 
  \phi_{{\rm new}} = \int^{\phi} d\phi
    \frac{L(\phi)^{1/2}}{|K(\phi)|^{1/2}} M_{s}^2,
\eeq
the function $p(\phi,\nabla\phi)$ has a simple form,
\beq
  p(\phi,\nabla\phi) = g(\phi) \lmk - X + \frac{X^2}{M_s^4} \rmk 
           + \sum_{i} h_{i}(\phi) \del_{\mu}\phi J_{i}^{\mu}
\eeq
with
\bea
  g(\phi) &=& \frac{K^2(\phi)}{L(\phi)} \frac{1}{M_s^4} \non \\
  h_{i}(\phi) &=& \frac{|K(\phi)|^{\frac12}}{L^{\frac12}(\phi)} 
                  \frac{M_{i}(\phi)}{M_s^2}.
\eea
Here the subscripts $new$ are omitted and the constant $M_s$ with the
dimension one is introduced to give the new field $\phi$ a correct
dimension. Thus, the quintessence field $\phi$ can have the derivative
coupling to the lepton current as given in the previous section after
replacing $h_i(\phi)$ by $c_{i} f(\phi) / M$.

In \cite{COY}, it is shown that, during the matter or radiation
dominated epoch, the scaling solution exists for the function $g(\phi)$
with the form of the inverse power law $g(\phi) = (\phi/M_s)^{-\alpha}
M_s^4$. So, we adopt such a function as the form of $g(\phi)$. Then, the
scaling solution is given by\footnote{It can be easily shown that the
modification of the equation of motion for $\phi$ due to the presence of
the derivative coupling is negligible in our context.}
\beq
 \phi = \xi M_s^2 t, \qquad \dot\phi = \xi M_s^2,
\eeq
where the coefficient $\xi = \sqrt{2(1-w_{Q})/(1-3w_{Q})}$, and $w_{Q} =
\rho_{Q}/p_{Q}$ is the equation of state for the quintessence field and
is given by
\beq
  w_Q = \frac{(1+w_B)\alpha}{2}-1.
\eeq
Here $w_B$ is the equation of state for the background matter or
radiation. Requiring that $w_Q<0$ during the matter dominated epoch,
the exponent $\alpha$ is constrained as $\alpha < 2$.

In order to fix the parameter $M_s$, we require that the energy density
of the quintessence field starts to dominate the energy density of the
Universe recently and obeys the scaling solution until recently, which
gives \cite{COY}
\beq
  M_s \sim 10^{\frac{43\alpha-48}{4-\alpha}} {\rm GeV}.
\eeq
Then, the value of the quintessence field at BBN is given by
\beq 
  \phi_{\rm BBN} 
       \sim 10^{\frac{62\alpha}{4-\alpha}} {\rm GeV}.
\eeq

Now, we discuss the spontaneous leptogenesis induced by the
quintessence field $\phi$. To make the discussion concrete, we set
$\alpha = 0.9$ as an example. In this case, $M_s \sim 10^{-3}$ GeV.
The discussion applies to the other cases in the same way except the
dependence on the parameters. Then, we assume that $f(\phi)$ has the
following form:
\beq
  f(\phi) = \lmk \frac{\phi M^5}{(\phi + M)^4 M_s^2} \rmk^{\frac12}.
\eeq
Such a coupling function may be obtained by the extension of the shift
symmetry introduced in \cite{KYY,YY}. This function $f(\phi)$ is
roughly classified into two regions,
\beq
  f(\phi) \simeq \left\{
             \begin{array}{ll}
              \lmk \frac{\phi M}{M_s^2} \rmk^{\frac12}, 
                 & \quad {\rm for} \quad \phi \lesssim M \\
              \lmk \frac{M^5}{\phi^3 M_s^2} \rmk^{\frac12},
                 & \quad {\rm for} \quad \phi \gtrsim M.
             \end{array}\right.
\eeq 
For $M = M_{G} \simeq 10^{18}$ GeV, we can use the second region for
the function $f(\phi)$ at least until BBN.

Using Eq. (\ref{eq:lepton}), the lepton asymmetry is estimated as
\beq
  \frac{n_{L}^{i}}{s} 
    \sim \frac{c_{i}}{M_G T} f(\phi)\, \dot\phi
    \sim c_{i} \lmk \frac{M_s}{T} \rmk^2
    \propto T^{-2}.
\eeq                             
By inserting $T_{\rm EW} \sim 100$ GeV and $M_s \sim 10^{-3}$ GeV into
the above equation, the total lepton asymmetry at the electroweak
phase transition is given by
\beq
  \left. \frac{n_{L}}{s} \right|_{T = T_{\rm EW}} 
    \sim c \,10^{-10}.
\eeq     
A part of the lepton asymmetry is changed into the baryon asymmetry
through the sphaleron effects \cite{sphaleron}. Thus, taking $c =
\sum_{i} c_{i} = - \CO(1)$, the present baryon asymmetry is given by
\beq
  \frac{n_B}{s} \sim 10^{-10}.
\eeq 

On the other hand, if the decoupling temperature $T_{D}$ of the lepton
number violating interaction is nearly equal or lower than $T_{\rm
  BBN} \sim 1$ MeV, the lepton asymmetry at BBN is given by
\beq
  \left. \frac{n_{L}^{i}}{s} \right|_{T = T_{\rm BBN}} 
    \sim c_{i}.
\eeq
Taking $c_{1} = \CO(1)$, the electron type asymmetry becomes positive
and of order unity at BBN.

Finally, we must check the present constraint on the derivative
coupling given in Eq. (\ref{eq:coupling}) with those from laboratory
experiments. For the time component, the coefficient $\mu(t_0)$ is
constrained as $|\mu(t_0)| \lesssim 10^{-25}$ GeV \cite{CKMP}. Here
$t_0$ is the present age of the Universe. In our model, $\mu(t_0)$ is
given by
\beq
  |\mu(t_0)| \sim \frac{|c|}{M} 
           \lmk \frac{M^5}{\phi_0^3 M_s^2} \rmk^{\frac12}
           \sim 10^{-32} {\rm GeV}
           \ll 10^{-25} {\rm GeV}
\eeq
for $|c|=\CO(1)$ and $M=M_{G}$.

\section{Discussion and conclusions}

\label{sec:con}

In this paper, we considered the large lepton asymmetry from a rolling
scalar field. Considering the derivative coupling of the scalar field
to the lepton number current, the presence of an effective nonzero
time derivative of the scalar field leads to $CPT$ violation, which
generates the lepton asymmetry even in thermal equilibrium. This
lepton asymmetry changes with time. So, depending on the dynamics of
the scalar field, it is possible that the lepton asymmetry is small at
the electroweak phase transition but large at BBN. A part of the
lepton asymmetry is converted into the baryon asymmetry with the
opposite sign through sphaleron effects. We pointed out that by
choosing the sign of the coupling constants properly, a large positive
lepton asymmetry of electron type and a small positive baryon
asymmetry can be realized simultaneously.

As an example, we considered the real rolling scalar field, which
realizes the k-essence model.  By considering the scaling solution, we
showed that this model manifests just such asymmetries. However, the
coupling function to the lepton number current is a bit complicated.
This is mainly because the time derivative of the scaling solution is
a constant irrespective of cosmic time.  If we abandon the scaling
solution, there may be evolution of the quintessence field, in which
the magnitude of the time derivative of the quintessence field at BBN
is much larger than that at the electroweak phase transition. For such
evolution, the model may work, in which the coupling to the lepton
number current is simple, that is, $f(\phi) \equiv 1$. Such a
possibility will be considered in a further publication.

Finally, we comment on the recent discussion of neutrino oscillations
around BBN. It was pointed out that complete or partial equilibrium
between all active neutrinos may be accomplished through neutrino
oscillations in the presence of neutrino chemical potentials,
depending on neutrino oscillation parameters \cite{equilibrium}. In
case of partial equilibrium, we need not change our scenario. Complete
equilibrium may spoil our scenario. In that case, if we choose the
coupling constants $c_{i}$ as $c_{1} = - c_{2} = - c_{3}$ for some
symmetry reason, our scenario still works. In such a case, no mixing
takes place because of the cancellation, as pointed out in
\cite{equilibrium}.

\subsection*{ACKNOWLEDGMENTS}

M.Y. is very grateful to W. Kelly for useful comments and correcting the
English. M.Y. is also grateful to T. Chiba and F. Takahashi for useful
comments. M.Y. is partially supported by a Japanese Grant-in-Aid for
Scientific Research from the Ministry of Education, Culture, Sports,
Science, and Technology.

\end{document}